\RequirePackage{fix-cm}
\documentclass[twocolumn,epjc3]{svjour3}  
\smartqed  
\RequirePackage{graphicx}
\RequirePackage{mathptmx}      

\usepackage{graphics}
\usepackage{subfiles}
\usepackage[english]{babel}
\usepackage{graphicx}
\usepackage{amssymb,amsmath,amstext,bm}
\usepackage{epsf,epsfig}
\usepackage{booktabs}
\usepackage{multirow}
\usepackage[utf8]{inputenc}
\usepackage{fancyhdr}
\usepackage{xcolor,color}
\journalname{Eur. Phys. J. A}
\begin{document}
\title{Deeply virtual Compton scattering on the neutron with positron beam}
\author{S. Niccolai\thanksref{e1,addr1}, P. Chatagnon\thanksref{addr1}, M. Hoballah\thanksref{addr1}, D. Marchand\thanksref{addr1}, C. Munoz Camacho\thanksref{addr1}, E. Voutier\thanksref{addr1}
}                     
%
%
\thankstext{e1}{e-mail: silvia@jlab.org}
\institute{Laboratoire de Physique des 2 Infinis Ir\`ene Joliot-Curie, Universit\'e Paris-Saclay, CNRS/IN2P3, France \label{addr1}}
\date{Received: date / Revised version: date}
%
\maketitle

\begin{abstract}
Measuring DVCS on a neutron target is a necessary step to deepen our understanding of the structure of the nucleon in terms of Generalized Parton Distributions (GPDs). The combination of neutron and proton targets allows to perform a flavor decomposition of the GPDs. Moreover, neutron-DVCS plays a complementary role to DVCS on a transversely polarized proton target in the determination of the GPD $E$, the least known and constrained GPD that enters Ji's angular momentum sum rule. A measurement of the beam-charge asymmetry (BCA) in the $e^{\pm} d\to e^{\pm}n\gamma(p)$ reaction can significantly impact the experimental determination of the real parts of the $E$ and, to a lesser extent, $\widetilde{H}$ GPDs.
\end{abstract}
\PACS{
      {PACS-key}{discribing text of that key}   \and
      {PACS-key}{discribing text of that key}
     } 

\section{Introduction}
It is well known that the fundamental particles which form hadronic matter are the quarks and the gluons, whose interactions are described by the QCD Lagrangian. However, exact QCD-based calculations cannot yet be performed to explain all the properties of hadrons in terms of their constituents. Phenomenological functions need to be used to connect experimental observables with the inner dynamics of the constituents of the nucleon, the partons. Typical examples of such functions include form factors, parton densities, and distribution amplitudes. The GPDs are nowadays the object of intense research in the perspective of unraveling nucleon structure. They describe the correlations between the longitudinal momentum and transverse spatial position of the partons inside the nucleon, they give access to the contribution of the orbital momentum of the quarks to the nucleon, and they are sensitive to the correlated $q$-$\bar{q}$ components. The original articles and general reviews on GPDs and details of the formalism can be found in Refs.~\cite{muller}-\cite{revrady}.

\begin{figure}[htb]
\begin{center}
\includegraphics[scale=0.47]{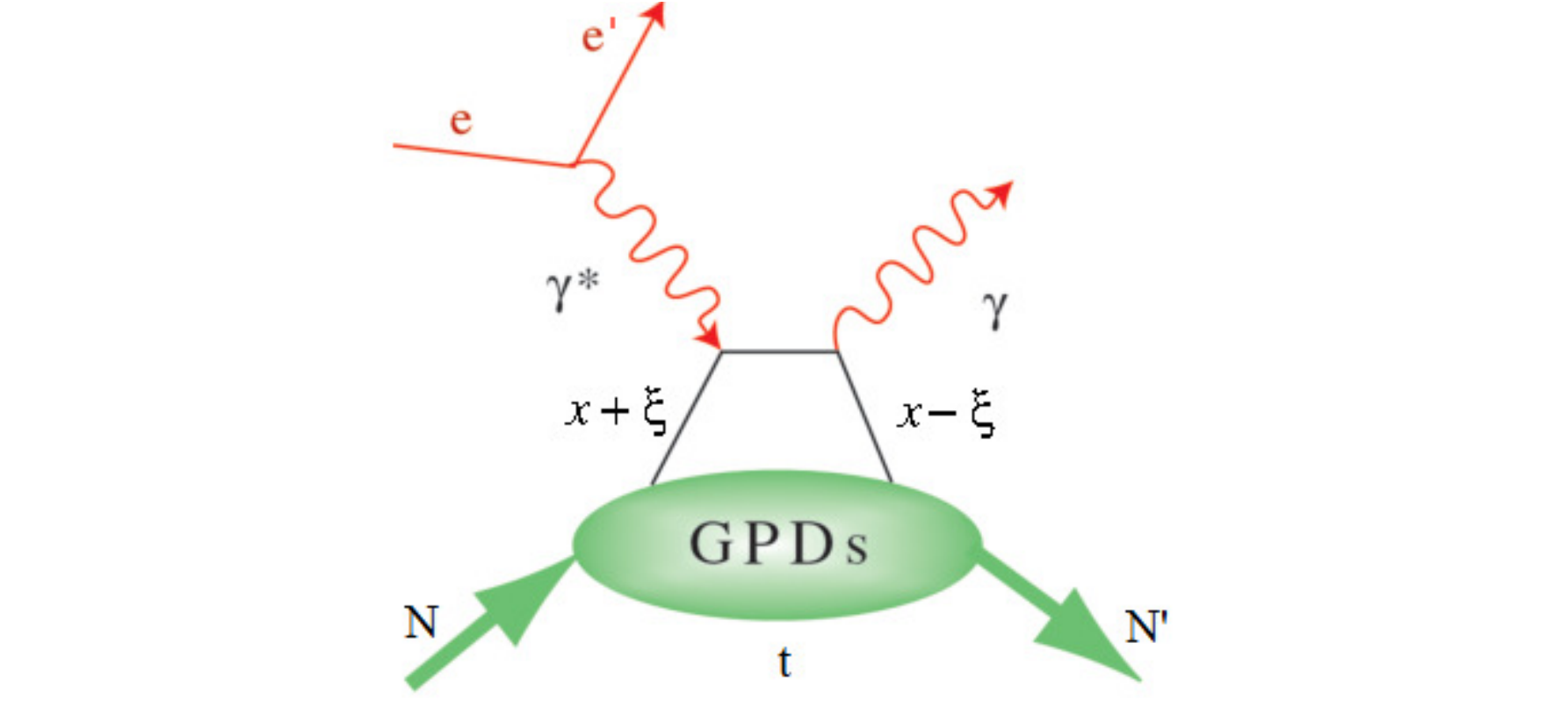}
\caption[Handbag diagram for the DVCS process] {The handbag diagram for the DVCS process on the nucleon $eN \to e'N'\gamma'$. Here $x+\xi$ and $x-\xi$ are the longitudinal momentum fractions of the struck quark before and after scattering, respectively, and $t=(N-N')^2$ is the squared four-momentum transfer between the initial and final nucleons. $\xi\simeq x_B/(2-x_B)$ is proportional to the Bjorken scaling variable $x_B=Q^2/2M\nu$, where $M$ is the nucleon mass and $\nu$ the energy transfer to the quark.}
\label{fig:dvcs}
\end{center}
\end{figure}
The GPDs of the nucleon are the structure functions which are accessed in the measurement of the exclusive leptoproduction of a photon (DVCS) or of a meson on the nucleon, at sufficiently large virtual-photon virtuality ($Q^2$) for the reaction to happen at the quark level. Figure~\ref{fig:dvcs} illustrates the leading process for DVCS, also called ``handbag diagram''. At leading-order QCD and at leading twist, considering only quark-helicity conserving quantities and the quark sector, the process is described by four GPDs: $H^q, \widetilde{H^q} , E^q, \widetilde{E^q}$, one for each quark flavor $q$, that account for the possible combinations of relative orientations of the nucleon spin and quark helicities between the initial and final states. $H^q$ and $E^q$ do not depend on the quark helicity and are therefore called unpolarized GPDs while $\widetilde{H^q}$ and $\widetilde{E^q}$ depend on the quark helicity and are called polarized GPDs. $H^q$ and $\widetilde{H^q}$ conserve the spin of the nucleon, whereas $E^q$ and $\widetilde{E^q}$ correspond to a nucleon-spin flip. 
\\
The GPDs depend upon three variables, $x$, $\xi$ and $t$: $x+\xi$ and $x-\xi$ are the longitudinal momentum fractions of the struck quark before and after scattering, respectively, and $t$ is the squared four-momentum transfer between the initial and final nucleon (see caption of Fig.~\ref{fig:dvcs} for definitions). The transverse component of $t$ is the Fourier-conjugate of the transverse position of the struck parton in the nucleon. Among these three variables, only $\xi$ and $t$ are experimentally accessible with DVCS. 
\\
The DVCS amplitude is proportional to combinations of integrals over $x$ of the form 
\begin{equation}\label{dvcs-ampl}
\int_{-1}^{1} d x \, F(\mp x,\xi,t)\left[\frac{1}{x-\xi+i\epsilon}\pm\frac{1}{x+\xi-i\epsilon}\right],
\end{equation}
where $F$ represents one of the four GPDs. The top combination of the plus and minus signs applies to unpolarized GPDs ($H^q, E^q$), and the bottom combination of signs applies to the polarized GPDs ($\widetilde{H^q}, \widetilde{E^q}$). Each of these 4 integrals, or Compton Form Factors (CFFs), can be decomposed into its real and imaginary parts, as following:
\begin{eqnarray}\label{def_cffs1}
\Re{\rm e} \left[{\cal F} (\xi,t)\right] & = & {\cal P}  \int_{-1}^{1}dx\left[\frac{F(x,\xi,t)}{x-\xi}\mp\frac{F(x,\xi,t)}{x+\xi}\right] \,\,\,\,\,\,\,\, \\
\Im{\rm m} \left[{\cal F} (\xi,t)\right] & = & -\pi [F(\xi,\xi,t)\mp F(-\xi,\xi,t)], \label{def_cffs2}
\end{eqnarray}
where ${\cal P}$ is Cauchy's principal value integral and the sign convention is the same as in Eq.~\ref{dvcs-ampl}. The information that can be extracted from the experimental data at a given ($\xi,t$) point depends on the measured observable. $\Re{\rm e} [{\cal F}]$ is accessed primarily measuring observables which are sensitive to the real part of the DVCS amplitude, such as double-spin asymmetries, beam-charge asymmetries or unpolarized cross sections. $\Im{\rm m} [{\cal F}]$ can be obtained measuring observables which are sensitive to the imaginary part of the DVCS amplitude, such as single-spin asymmetries or the difference of polarized cross-sections. However, knowing the CFFs does not define the GPDs uniquely. A model input is necessary to deconvolute their $x$-dependence. 
The DVCS process is accompanied by the Bethe-Heitler (BH) process, in which the final-state real photon is radiated by the incoming or scattered electron and not by the nucleon itself. The BH process, which is not sensitive to GPDs, is experimentally indistinguishable from DVCS and interferes with it at the amplitude level. However, considering that the nucleon form factors are well known at small $t$, the BH process is precisely calculable.
\section*{Neutron GPDs and flavor separation}
The importance of neutron targets in the DVCS phenomenology was clearly established in the pioneering Hall A experiments at 6 GeV, where the polarized-beam cross section difference off a neutron, from a deuterium target, was measured~\cite{malek,benali}.  Measuring neutron GPDs in complement to proton GPDs allows for a quark-flavor separation. For instance, the ${\cal E}$-CFF of the proton and the neutron can be expressed as 
\begin{eqnarray}
{\cal E}_p(\xi, t) & = & \frac{4}{9}{\cal E}^u(\xi, t)+\frac{1}{9}{\cal E}^d(\xi, t) \\
{\cal E}_n(\xi, t) & = & \frac{1}{9}{\cal E}^u(\xi, t)+\frac{4}{9}{\cal E}^d(\xi, t)  
\end{eqnarray}
(and similarly for ${\cal H}$, ${\widetilde {\cal H}}$ and ${\widetilde {\cal E}}$). From this it follows that
\begin{eqnarray}
{\cal E}^u(\xi, t)=\frac{9}{15} \left[4 {\cal E}_p(\xi, t)-{\cal E}_n(\xi, t)\right] \\
{\cal E}^d(\xi, t)=\frac{9}{15} \left[4 {\cal E}_n(\xi, t)-{\cal E}_p(\xi, t)\right] \, .  
\end{eqnarray}
An extensive experimental program dedicated to the measurement of the DVCS reaction on a proton target has recently started at Jefferson Lab, in particular with CLAS12. Single-spin asymmetries with polarized beam and/or linearly or transversely polarized proton targets, as well as unpolarized and polarized cross sections, will be measured with high precision over a vast kinematic coverage. If a similar program is performed on the neutron, the flavor separation of the various GPDs will be possible. The beam-spin asymmetry for nDVCS, particularly sensitive to the GPD $E_n$ is currently being measured at CLAS12, using an experimental technique different from the initial Hall A measurement and involving the detection of the active neutron \cite{proposal}. Additionally, the measurement of single- and double-spin asymmetries with a longitudinally polarized deuteron target is also foreseen for the nearby future with CLAS12. Here we focus on the extraction of one more observable, the beam-charge asymmetry. The next section outlines the benefits of this observable for the CFFs determination. 

\section*{Beam charge asymmetry}
Considering unpolarized electron and positron beams, the sensitivity of the cross-section to the lepton beam charge can be expressed with the beam charge asymmetry observable 
\begin{equation}
A_{\rm C}(\phi) = \frac{d^4\sigma^+ - d^4\sigma^-} {d^4\sigma^+ + d^4\sigma^-} =  
                  \frac{d^4\sigma_{UU}^{\rm I}} {d^4\sigma_{UU}^{\rm BH} + d^4\sigma_{UU}^{\rm DVCS}} \label{eq:AC}
\end{equation}
which isolates the BH-DVCS interference contribution at the numerator and the DVCS amplitude at the denominator. Following the harmonic decomposition proposed in Ref.~\cite{belitski},
\begin{eqnarray}
d^4 \sigma_{UU}^{\rm I} & = & \frac{K_1}{\mathcal{P}_1(\phi) \, \mathcal{P}_2(\phi)} \sum_{n=0}^3 c_{n, \rm {unp}}^{\rm I}\cos(n\phi) \label{eq:int} \\
d^4 \sigma_{UU}^{\rm DVCS} & = & \frac{K_2}{Q^2} \sum_{n=0}^2 c_{n, \rm {unp}}^{\rm DVCS}\cos(n\phi) ,
\end{eqnarray}
where $K_i$'s are kinematical factors, and $P_i(\phi)$'s are the BH propagators. Because of the $1/Q^2$ kinematical suppression of the DVCS amplitude, the dominant contribution to the denominator originates from the BH amplitude. This approximation depends on the kinematics and, in the most general case, the DVCS contribution in the denominator complicates the extraction of CFFs. At leading twist, the dominant coefficients of the numerator are $c_{0,{\rm{unp}}}^{\rm I}$ and $c_{1,{\rm{unp}}}^{\rm I}$  
\begin{eqnarray}
c_{0,\rm{unp}}^{\rm I} & \propto & - \frac{\sqrt{-t}}{Q} \, c_{1,\rm{unp}}^{\rm I} \label{eq:r} \\
\label{eq:cosphi_term_bca}
c_{1,{\rm{unp}}}^{\rm I} & \propto & \Re{\rm e} \left[F_1 { \cal H}+\xi (F_1 + F_2)\widetilde{\cal H}-\frac{t}{4M^2}F_2{\cal E} \right] \, .
\end{eqnarray}
Given the relative strength of $F_1$ and $F_2$ at small $t$ for a neutron target, the beam charge asymmetry becomes
\begin{equation}
A_{\rm C}(\phi) \propto \frac{1}{F_2} \, \Re{\rm e} \left[ \xi \widetilde{\cal H}_n - \frac{t}{4M^2} {\cal E}_n \right] . 
\end{equation}
Therefore, the BCA is mainly sensitive to the real part of the GPD $E_n$, and, for selected kinematics, to the real part of the GPD $\widetilde{H_n}$. 

\section{Projections for CLAS12 experiment}\label{sec_countrate}

\begin{figure}
\begin{center}
\includegraphics[width=0.4\textwidth]{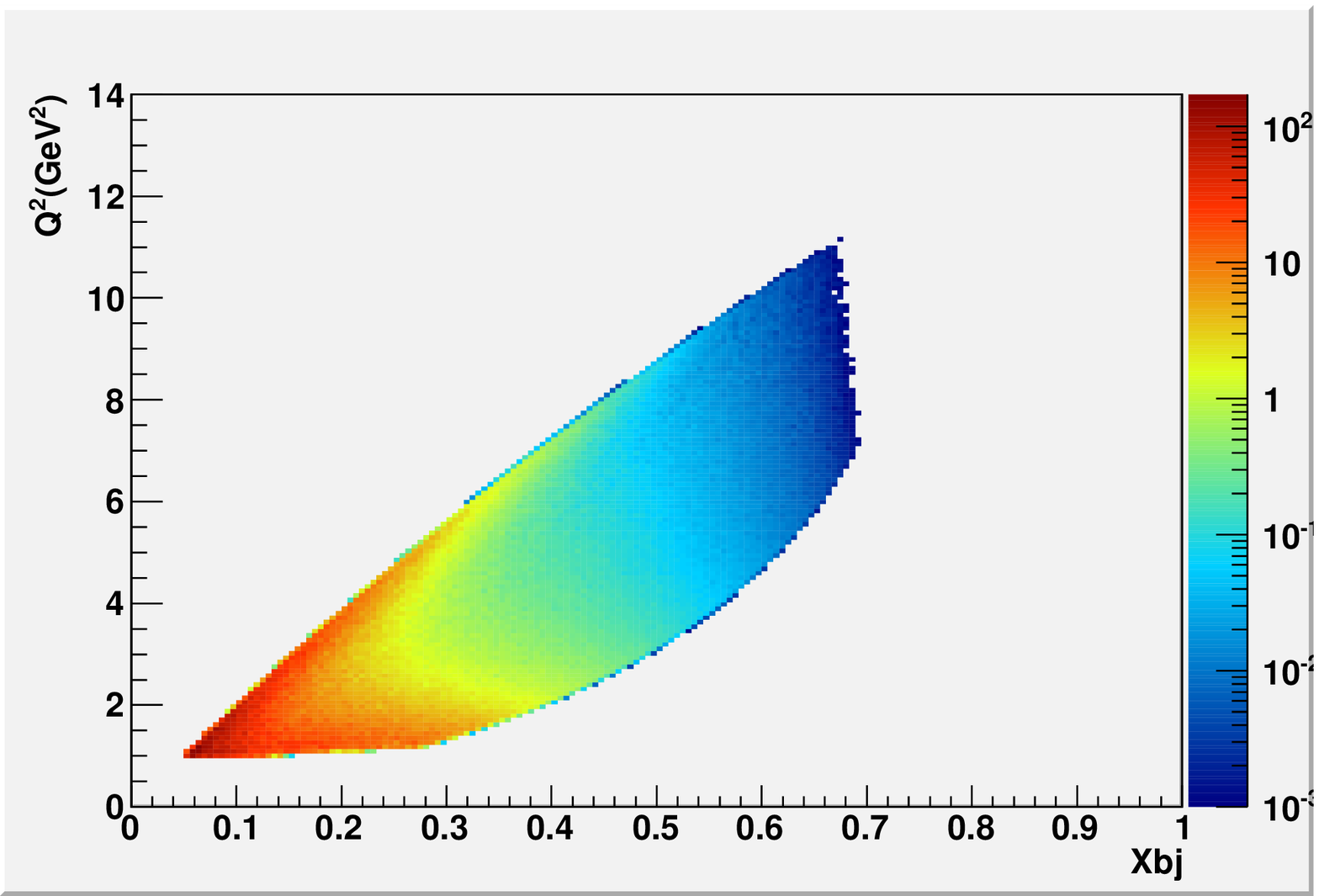}
\includegraphics[width=0.4\textwidth]{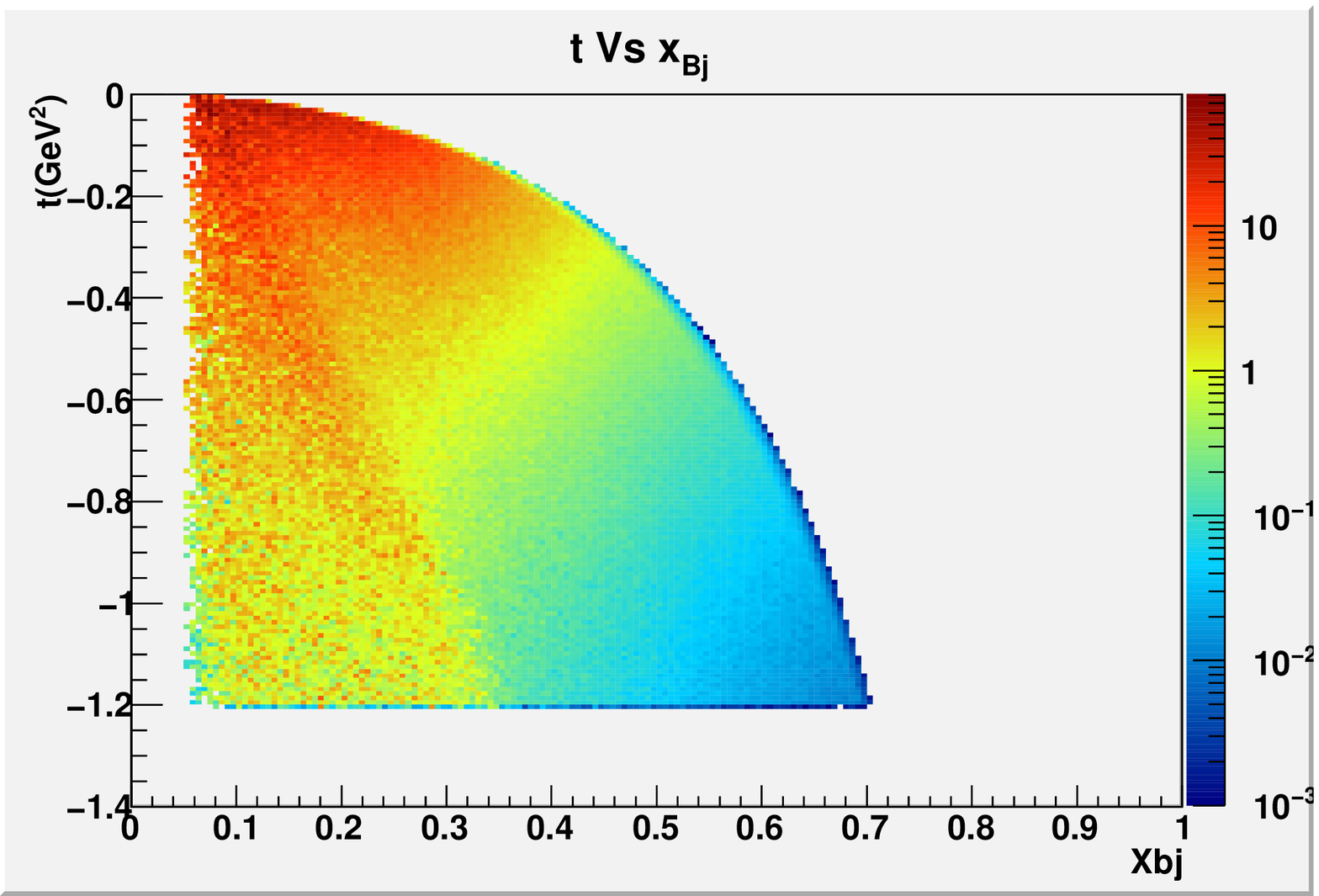}
\includegraphics[width=0.4\textwidth]{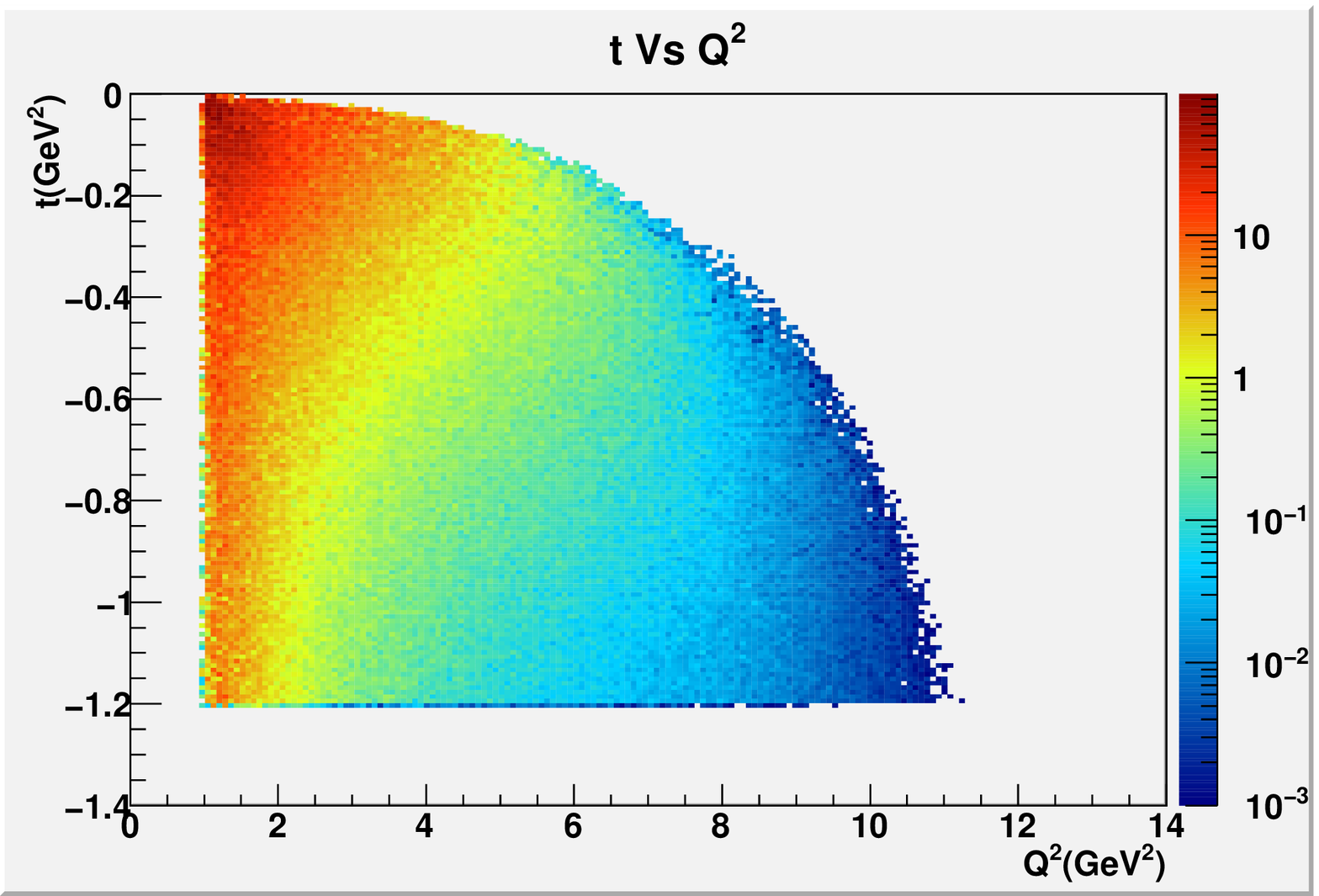}
\caption {Distributions of kinematic variables for n-DVCS events. Forward-CLAS12 acceptance cuts are included. Top: $Q^2$ as a function \
of $x_B$. Middle: $t$ as a function of $x_B$. Bottom: $t$ as a function of $Q^2$.}
\label{kine_vars}
\end{center}
\end{figure}

Projections for a neutron-DVCS experiment with positron and electron beams and the JLab CLAS12 spectrometer were produced. 
The n-DVCS/BH final state will be reconstructed by detecting the scattered electron/positron and the DVCS/BH photon in the forward part of CLAS12 and the recoil neutron mostly in the Central Neutron Detector (CND), as very few neutrons are emitted in the forward direction with enough momentum to be detected in EC with appreciable efficiency.
The expected number of reconstructed events for n-DVCS/BH has been calculated, as a function of the kinematics, with a GPD-based event generator including Fermi motion for a deuterium target and the Fast-MC simulation of CLAS12. The forward-CLAS12 fiducial cuts have been included, and an overall 10\% neutron-detection efficiency for neutrons with $\theta>40^o$ has 
been assumed. 
The detection efficiencies for electrons/positrons and photons efficiencies for the Forward Detector have been assumed to be 100\%, within the fiducial cuts. 
Figure~\ref{kine_vars} shows the coverage in $Q^2$, $x_B$ and $t$ that is obtained with an electron/positron beam energy of 11 GeV.
The count-rate calculation has been done for a luminosity $L=10^{35}$ cm$^{-2}$s$^{-1}$ per nucleon (corresponding to roughly 60 nA of beam current) and for 80 days (40 with positron beam, 40 with electron beam) of running time. A 4-dimensional ($Q^2$, $x_B$, $-t$, $\phi$) grid of bins has been adopted.

The BCA is defined, for each 4-dimensional bin, as:
\begin{eqnarray}
A=\frac{d^4\sigma^+ - d^4\sigma^-}{d^4\sigma^+ + d^4\sigma^-}=\frac{\frac{N^+}{Q^+}-\frac{N^-}{Q^-}}{\frac{N^+}{Q^+}+\frac{N^-}{Q^-}},
\end{eqnarray}

where $Q^{\pm}$ is the integrated charge for events with positron and electron beam. Here we assumed $Q^+=Q^-$.

The number of events $N^{\pm}$, for each 4-dimensional bin ($Q^2$, $x_B$, $t$ and $\phi$), has been computed as:

\begin{eqnarray}
N = \frac{d\sigma}{dQ^2dx_Bdtd\phi}\cdot \Delta t\cdot \Delta Q^2 \cdot\Delta x_B\cdot \Delta\phi\cdot L\cdot T\cdot A,
\end{eqnarray}

 where $\frac{d\sigma}{dQ^2dx_Bdtd\phi}$ is the 4-fold differential cross section, $T$ is the running time, $L$ the luminosity, $Acc$ is the bin-by-bin acceptance multiplied by the neutron-detection efficiency.

\begin{figure*}[hbt]
\begin{center}
\includegraphics[width=0.8\textwidth]{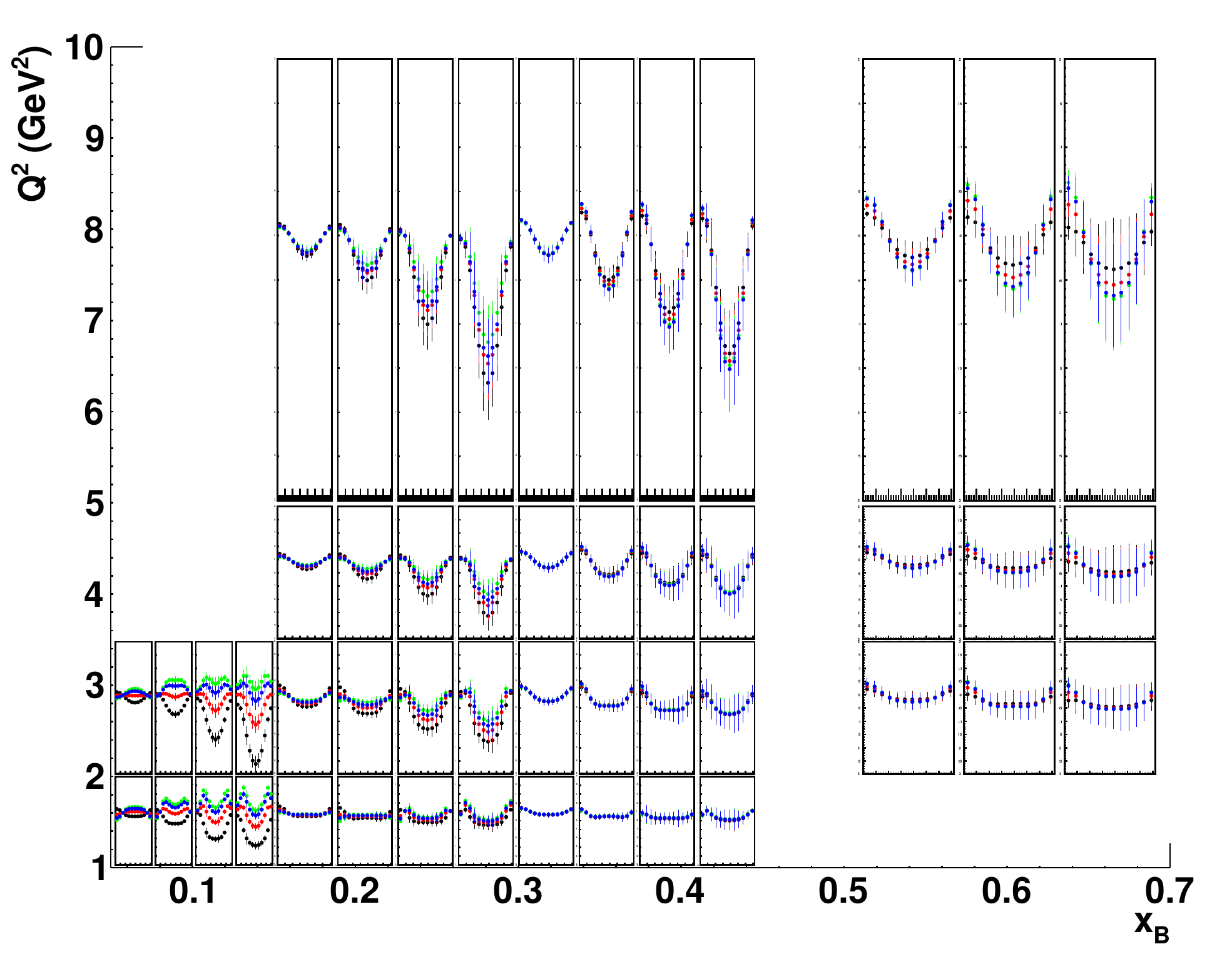}
\caption {Beam-charge asymmetry for n-DVCS/BH as predicted by the VGG model. Projections for different sets of values of $J_u$/$J_d$ are compared: 0.3/0.1 (black), 0.2/0.0 (red), 0.1/-0.1 (green), 0.3/-0.1 (blue)), plotted as a function of $\phi$ for all the\
 covered bins in $Q^2$, $x_B$, and $-t$. The vertical axis scale ranges from -0.3 to 0.1 for the top plot and from -0.3 to 0.2 for the bottom plot. The error \
bars reflect the expected uncertainties for our CLAS12 experiment, corresponding to 80 hours of beam time at a luminosity of $10^{35}$ cm$^{-2}$s$^{-1}$ per n\
ucleon.}
\label{asym_ndvcs}
\end{center}
\end{figure*}

The statistical errors on the beam-charge asymmetries will then depend on the values of the BCA itself ($A$), through the formula:
\begin{eqnarray}
\sigma_A = \frac{\sqrt{1-A^2}}{\sqrt{N}}.
\end{eqnarray}
Figure~\ref{asym_ndvcs} shows the expected accuracy on the n-DVCS/BH beam-charge asymmetry, computed using the VGG model and assuming various values for $J_u=.3$ and $J_d=.1$, for all 4-dimensional kinematic bins within the CLAS12 acceptance.

\section*{Extraction of Compton form factors} \label{sec_cff}
\begin{figure}[htb]
\begin{center}
\includegraphics[width=0.48\textwidth]{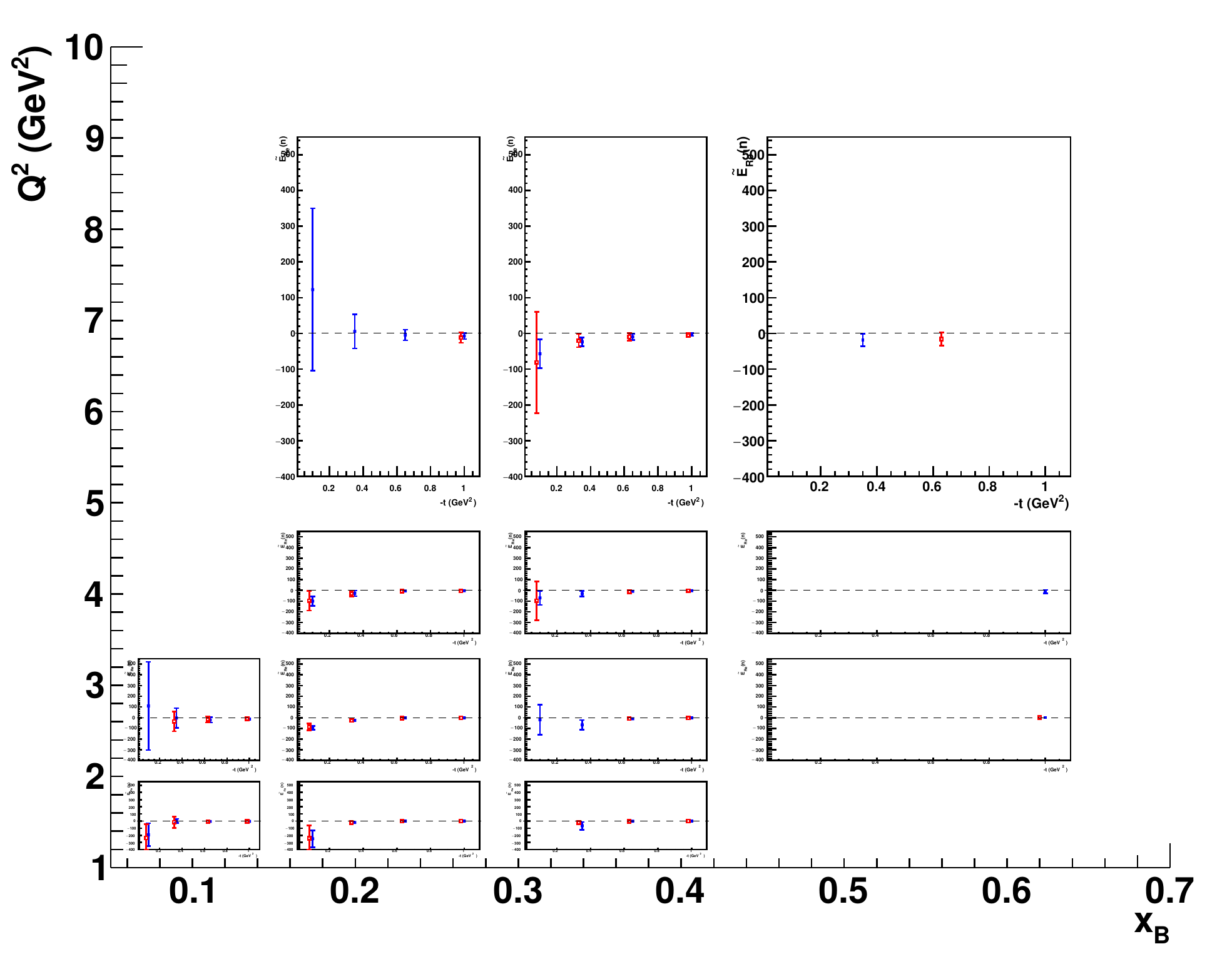}
\caption{$\tilde{E}_{Re}(n)$ as a function of $-t$, for all bins in $Q^2$ and $x_B$. The blue points are the results of the fits including the proposed BCA while the red ones include only already approved experiments.}
\label{cff_etre}
\end{center}
\end{figure}

\begin{figure}[htb]  
\begin{center}
\includegraphics[width=0.48\textwidth]{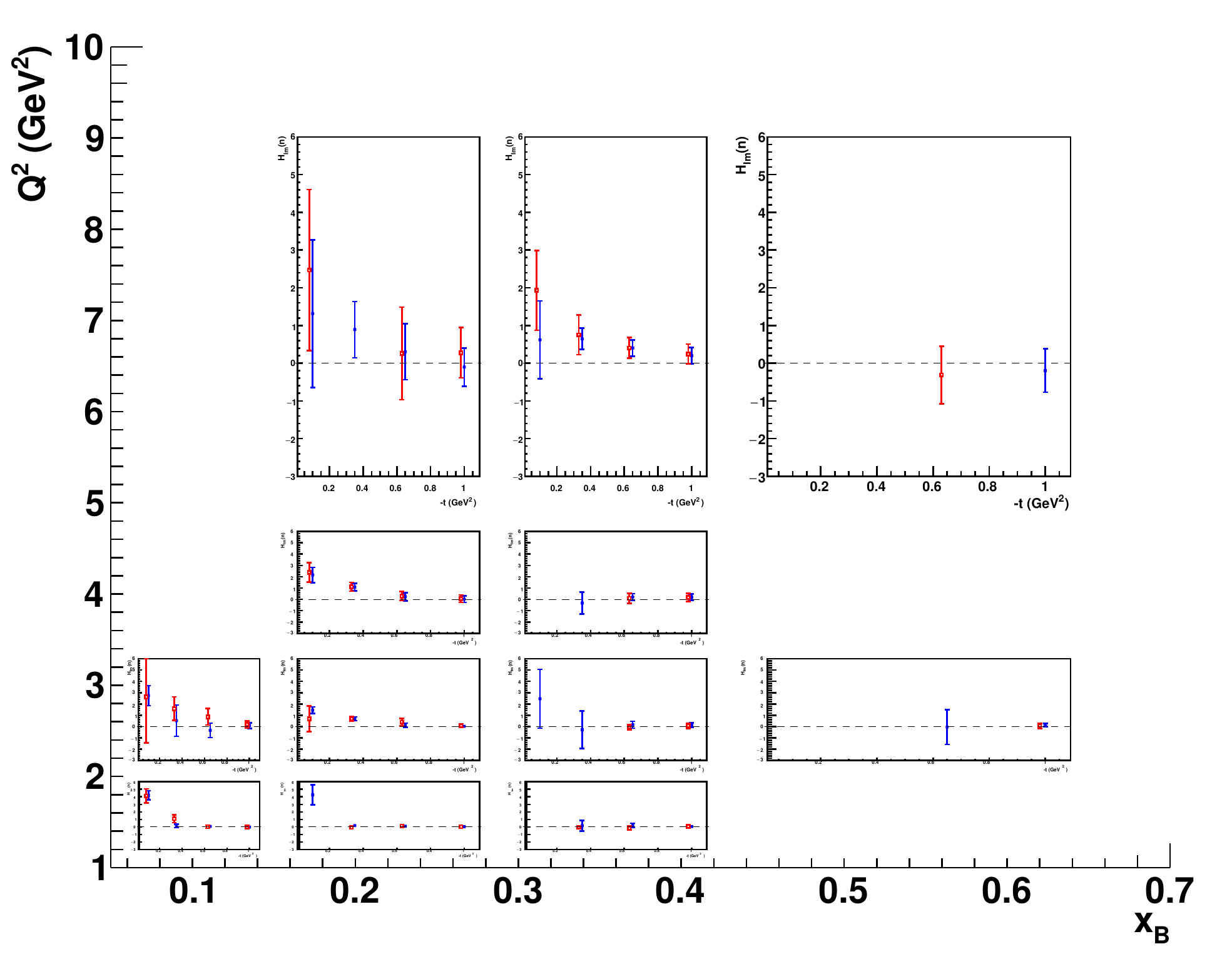}
\caption{$H_{Im}(n)$ as a function of $-t$, for all bins in $Q^2$ and $x_B$. The blue points are the results of the fits including the proposed BCA while the red ones include only already approved experiments.}
\label{cff_him}
\end{center}
\end{figure}

In order to establish the impact of a beam-charge asymmetry measurement on the nDVCS experimental program planned with CLAS12 at Jefferson Lab,  projections for four kinds of asymmetries (beam-spin asymmetry, BSA, longitudinal single, TSA, and double targe-spin asymmetry, DSA, and the BCA) were produced using the VGG model and realistic count rates and acceptances. The projected observables were then processed using a fitting procedure~\cite{fitmick,mick_herve} to extract the neutron CFFs. This approach is based on a local-fitting method at each given experimental $(Q^2, x_B,-t)$ kinematic point. In this framework, there are eight real CFF-related quantities 
\begin{eqnarray}
F_{Re}(\xi,t) & = & \Re{\rm e} \left[ {\cal F}(\xi,t) \right] \\
F_{Im}(\xi,t) & = & -\frac{1}{\pi} \Im{\rm m} \left[ {\cal F}(\xi,t) \right] \nonumber \\ 
& = & \left[ F(\xi,\xi,t)\mp F(-\xi,\xi,t) \right], 
\end{eqnarray}
where the sign convention is the same as for Eq.~\ref{dvcs-ampl}. These CFFs are the almost-free\footnote{The values of the CFFs are allowed to vary within $\pm 5$ times the values predicted by the VGG model~\cite{vgg1,vgg2}}  parameters to be extracted from DVCS observables using the well-established theoretical description of the process based on the DVCS and BH mechanisms. The BH amplitude is calculated exactly while the DVCS one is determined at the QCD leading twist~\cite{vgg1}. 
As there are eight CFF-related free parameters, including more observables, measured at the same kinematic points, will result in tighter constraint on the fit and will increase the number and accuracy of the extracted CFFs. In the adopted version of the fitter code, $\widetilde{E_{Im}}(n)$ is set to zero, as $\widetilde{E_n}$ is assumed to be purely real. Thus, seven out of the eight real and imaginary parts of the CFFs are left as free parameters in the fit. The results for the 7 neutron CFFs are shown in Figs.~\ref{cff_him}-\ref{cff_etre}, as a function of $-t$, and for each bin in $Q^2$ and $x_B$. The blue points are the CFFs resulting  from the fits of the four observables, while the red ones are the CFFs obtained fitting only the projections of the currently approved n-DVCS experiments. The error bars reflect both the statistical precision of the fitted observables and their sensitivity to that particular CFFs. Only results for which the error bars are non zero, and therefore the fits properly converged, are included in the figures. \newline
The major impact of the BCA measurement is, as expected, on $E_{Re}(n)$, for which the already approved projections have hardly any sensitivity. Thanks to the BCA, $E_{Re}(n)$ can be extracted over basically the whole phase. A considerable extension in the coverage can be also obtained for $\tilde{H}_{Re}(n)$. An overall improvement to the precision on the other CFFs, as well as an extension in their kinematic coverage can also be induced by the nDVCS BCA dataset. 

\begin{figure}[htb]  
\begin{center}
\includegraphics[width=0.48\textwidth]{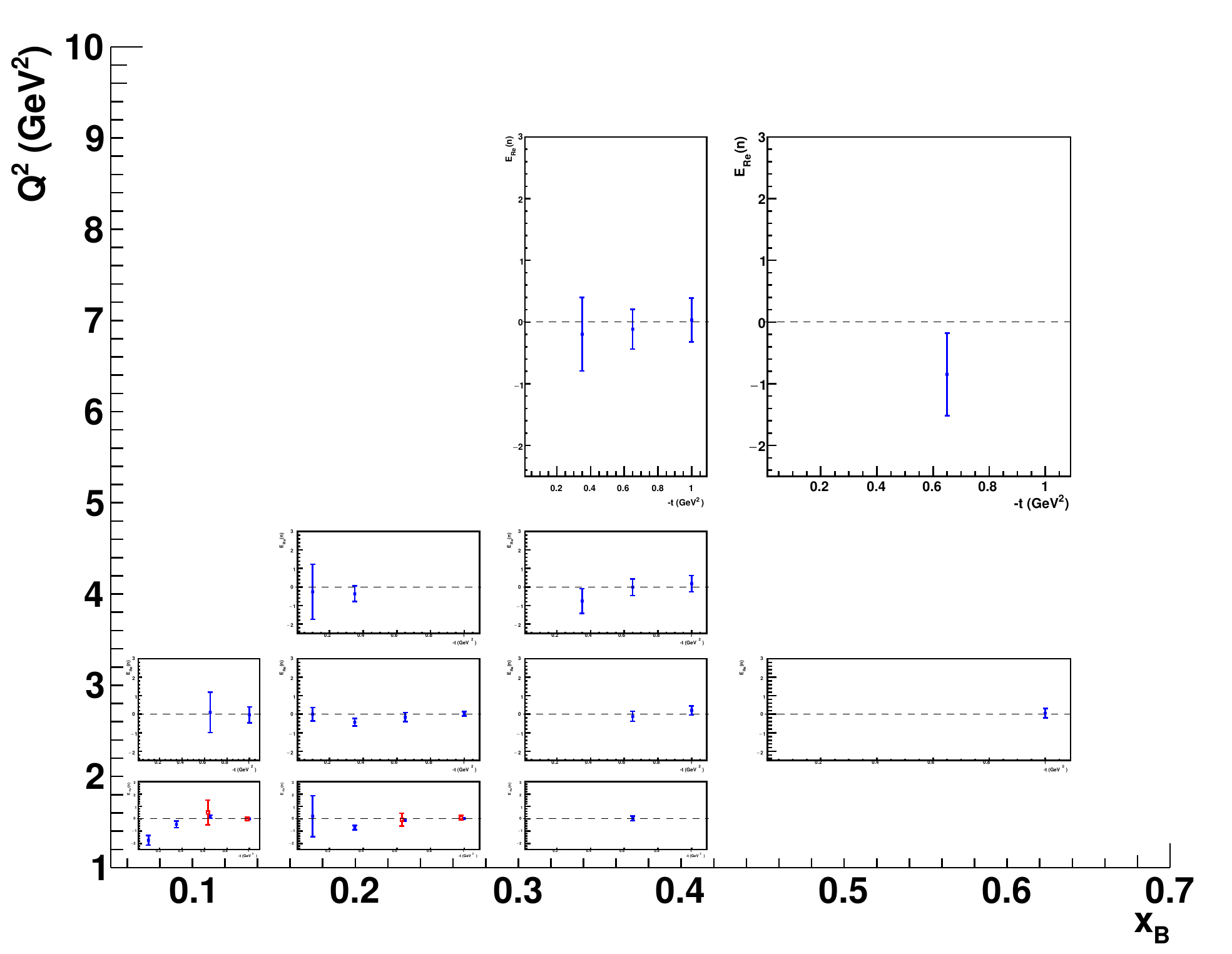}
\caption{$E_{Re}(n)$ as a function of $-t$, for all bins in $Q^2$ and $x_B$. The blue points are the results of the fits including the proposed BCA while the red ones include only already approved experiments.}
\label{cff_ere}
\end{center}
\end{figure}
\begin{figure}[htb]
\begin{center}
\includegraphics[width=0.48\textwidth]{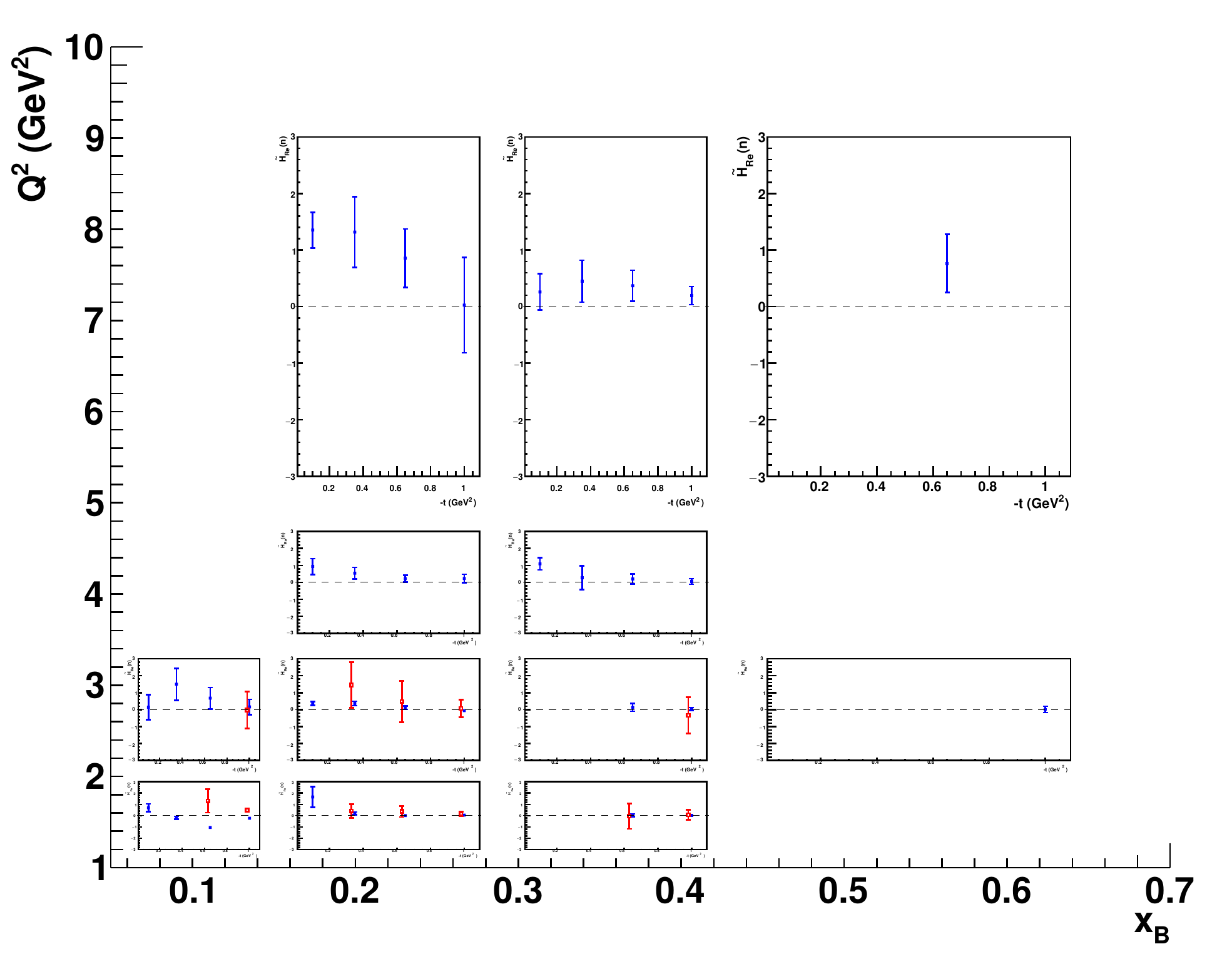}
\caption{$\tilde{H}_{Re}(n)$ as a function of $-t$, for all bins in $Q^2$ and $x_B$. The blue points are the results of the fits including the proposed BCA while the red ones include only already approved experiments.}
\label{cff_htre}
\end{center}
\end{figure}
\begin{figure}[htb]
\begin{center}
\includegraphics[width=0.48\textwidth]{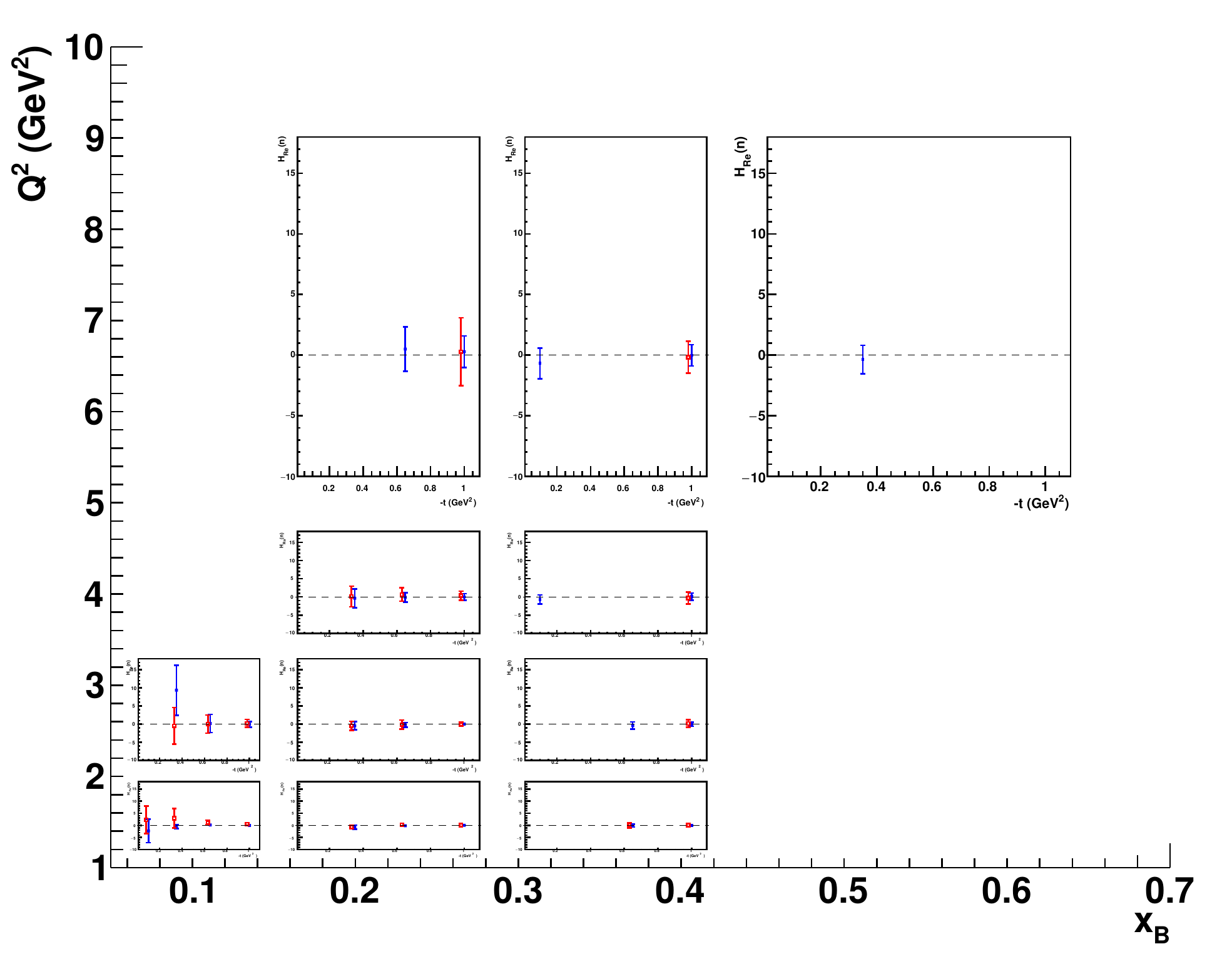}
\caption{$H_{Re}(n)$ as a function of $-t$, for all bins in $Q^2$ and $x_B$. The blue points are the results of the fits including the proposed BCA while the red ones include only already approved experiments.}
\label{cff_hre}
\end{center}
\end{figure}
\begin{figure}[htb]
\begin{center}
\includegraphics[width=0.48\textwidth]{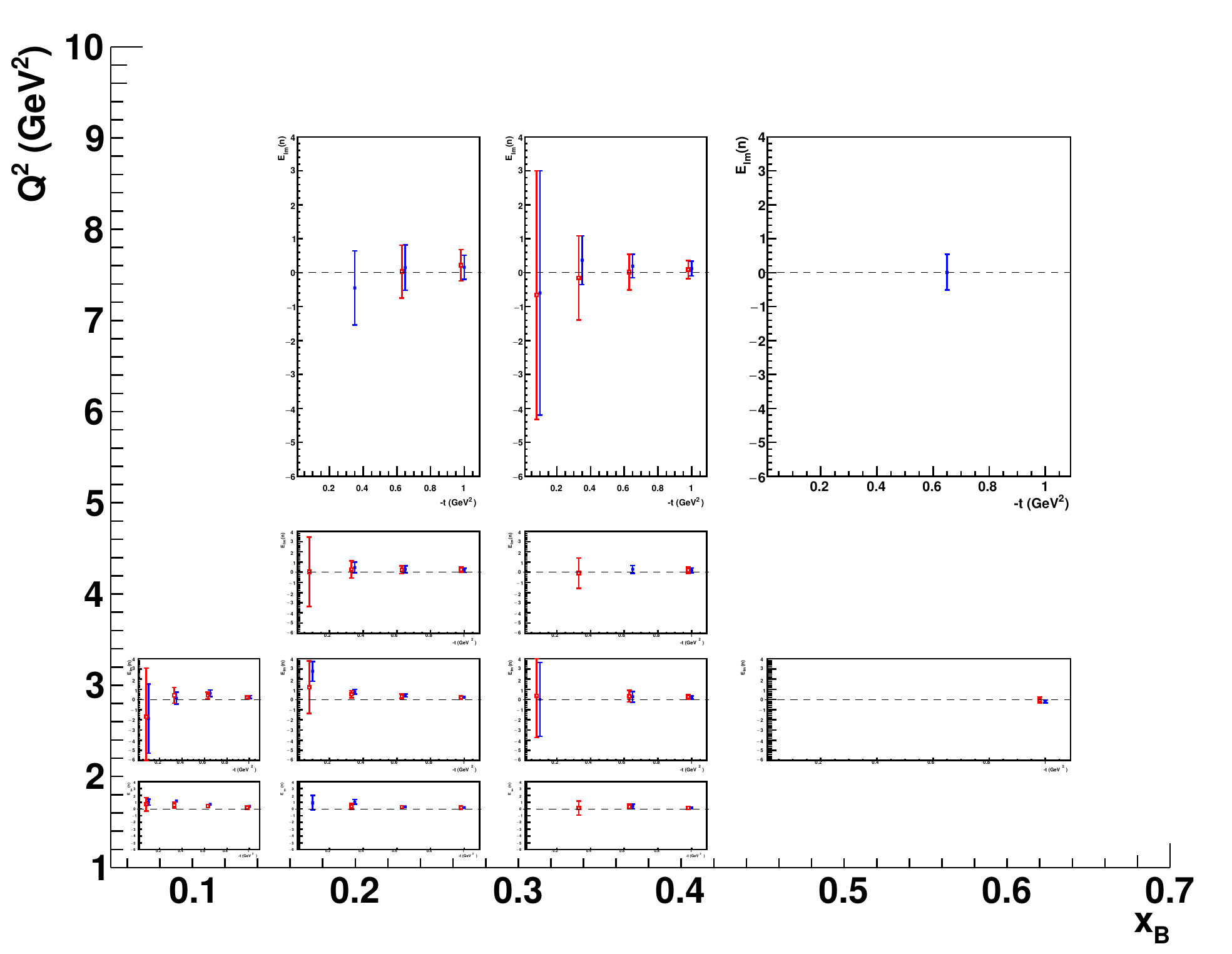}
\caption{$E_{Im}(n)$ as a function of $-t$, for all bins in $Q^2$ and $x_B$. The blue points are the results of the fits including the proposed BCA while the red ones include only already approved experiments.}
\label{cff_eim}
\end{center}
\end{figure}
\begin{figure}[htb]  
\begin{center}
\includegraphics[width=0.48\textwidth]{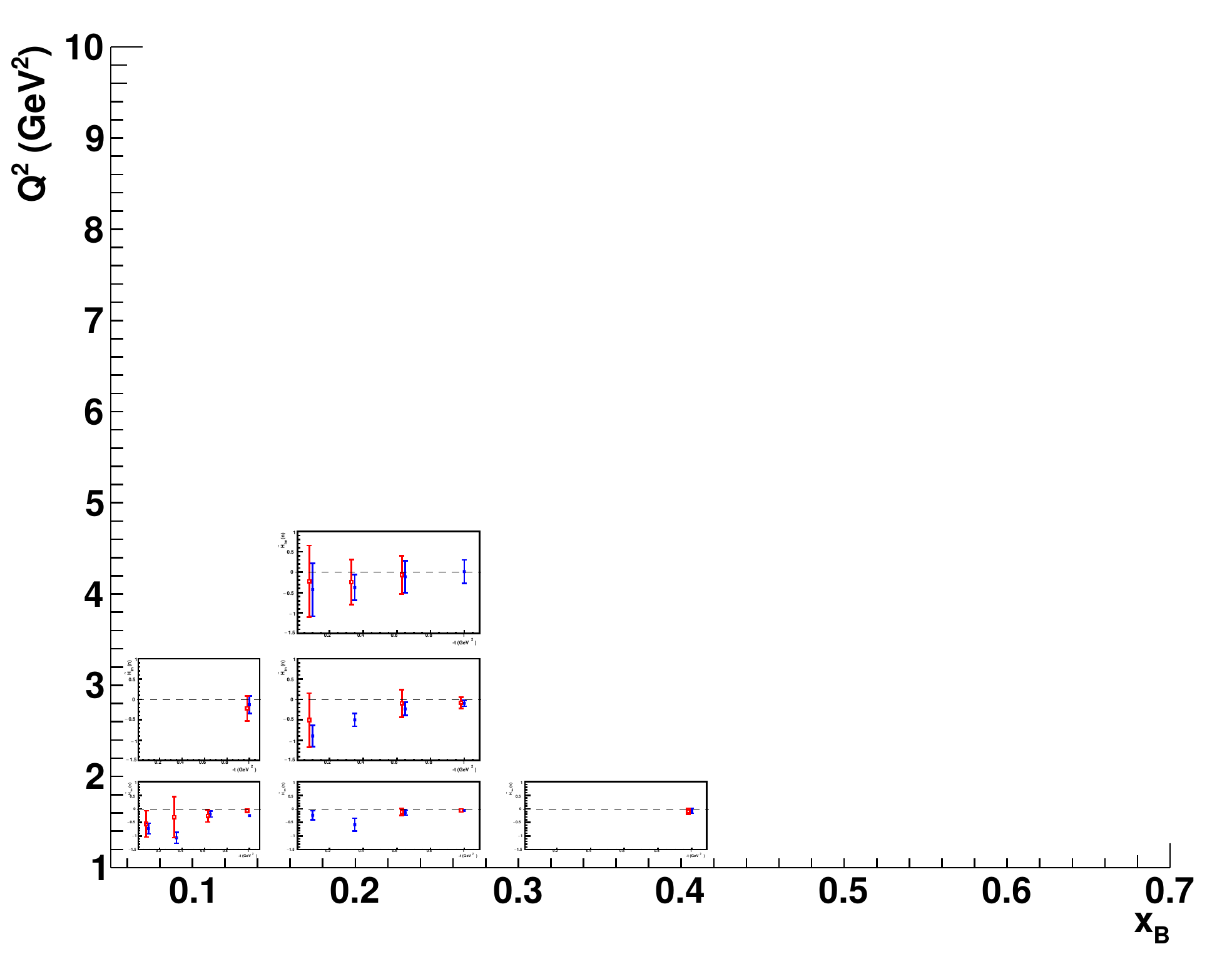}
\caption{$\tilde{H}_{Im}(n)$ as a function of $-t$, for all bins in $Q^2$ and $x_B$. The blue points are the results of the fits including the proposed BCA while the red ones include only already approved experiments.}
\label{cff_htim}
\end{center}
\end{figure}

\section*{Summary}
The strong sensitivity to the real part of the GPD $E^q$ of the beam-charge asymmetry for DVCS on a neutron target makes the measurement of this observable particularly important for the experimental GPD program of Jefferson Lab. This sensitivity is maximal for values of $x_B$ which are attainable only with a 11~GeV beam. Model predictions show that for possible CLAS12 kinematics, this asymmetry can be comparable in size to the BSA obtained for p-DVCS. \newline
The addition of the beam-charge asymmetry to the already planned measurements with CLAS12, permits the model-independent extraction of the real parts of the ${\cal E}_n$ and $\widetilde{{\cal H}_n}$ CFFs of the neutron over the whole available phase space. Combining all the neutron and the proton CFFs, obtained from the fit of n-DVCS and p-DVCS observables to be measured at CLAS12, will ultimately allow the quark-flavor separation of all GPDs.

\end{document}